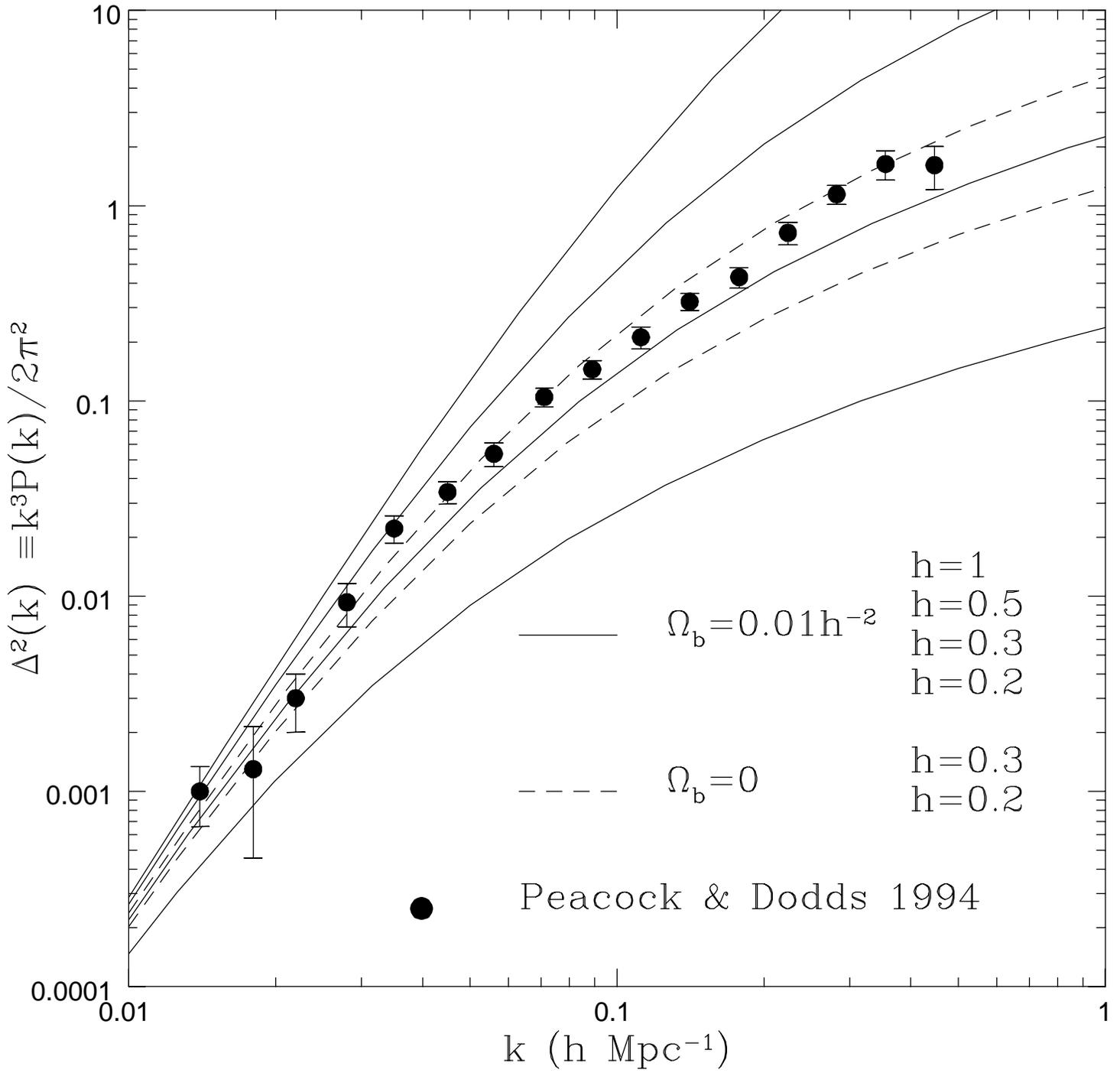

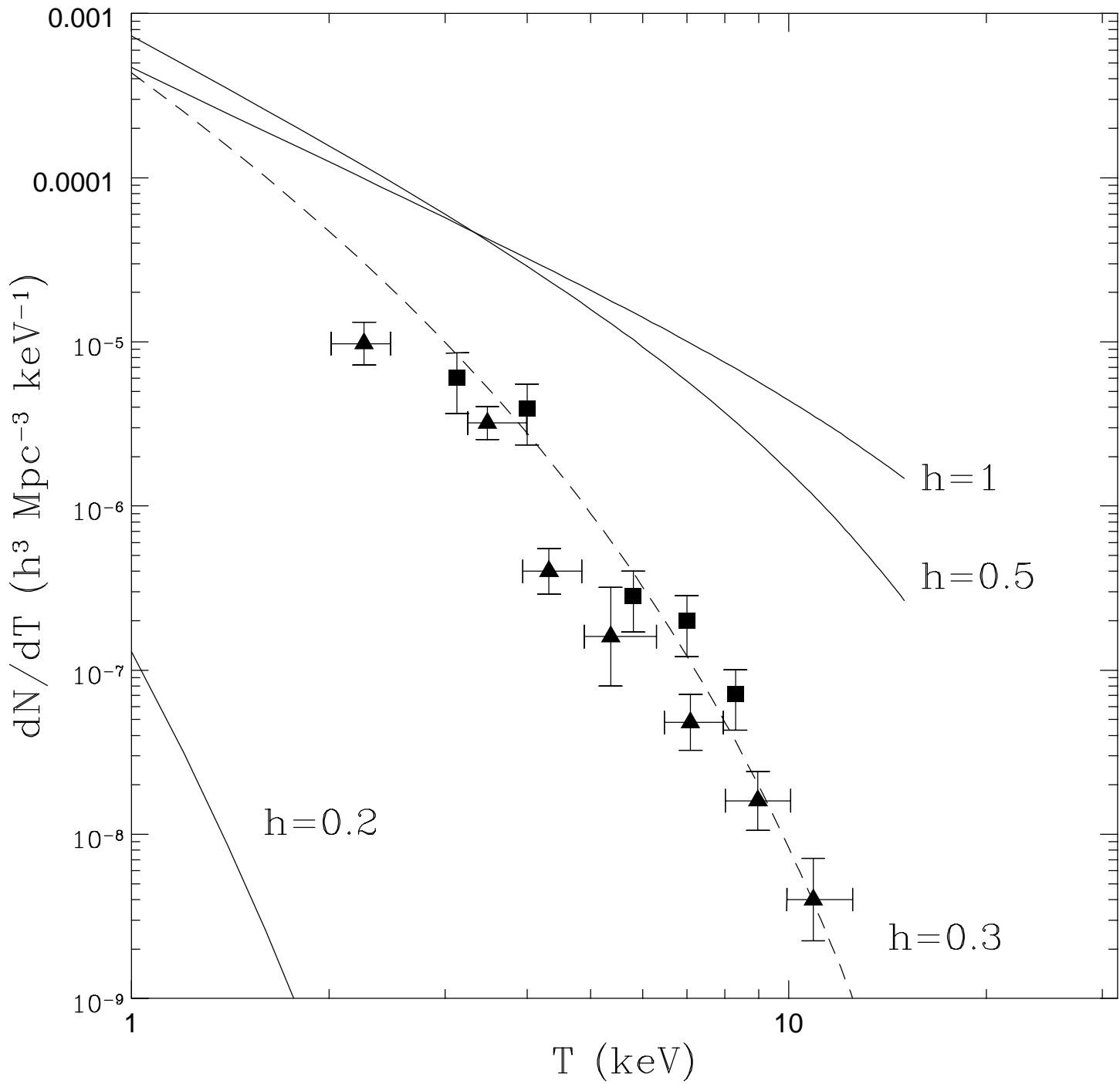

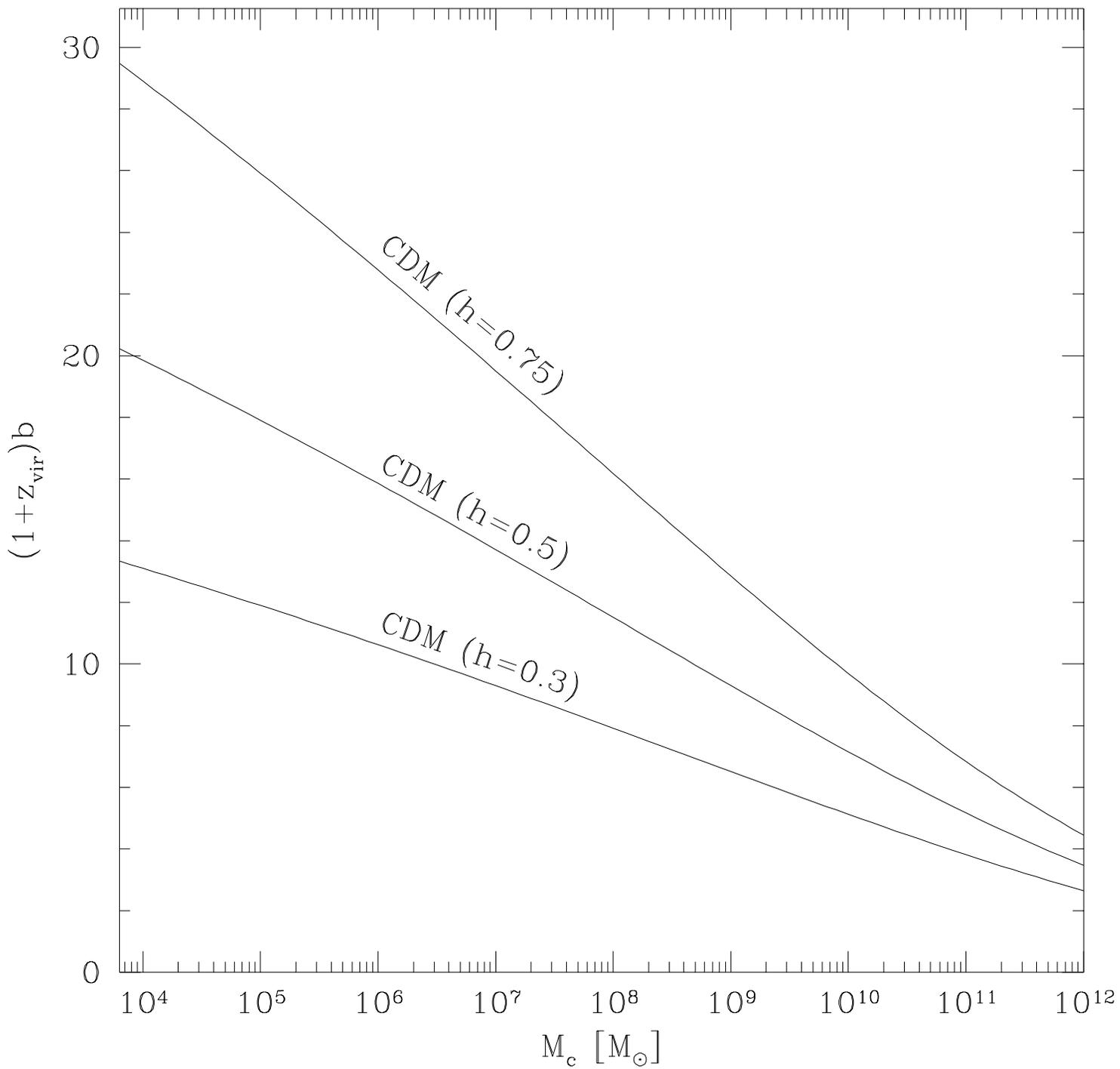

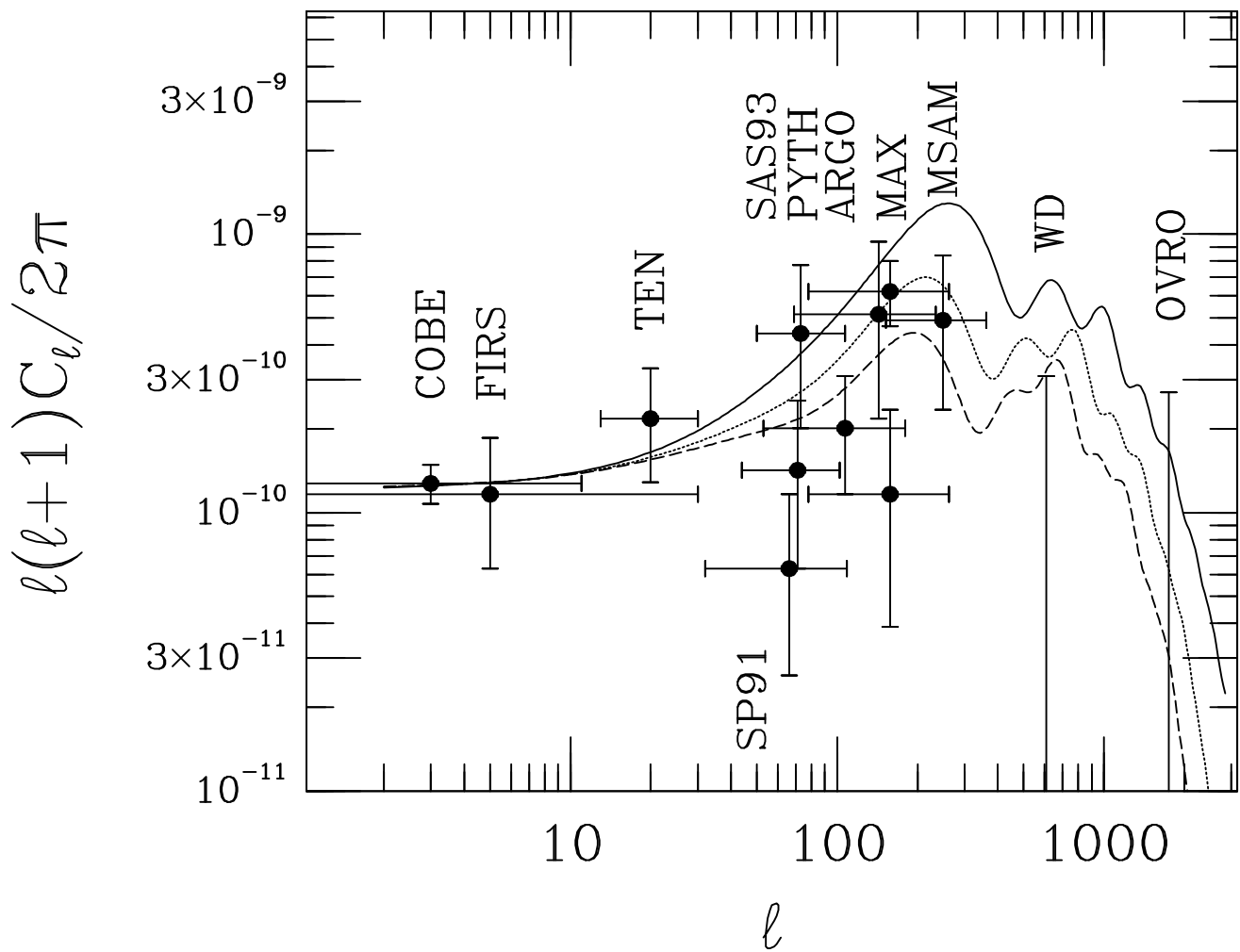

# THE CASE FOR A HUBBLE CONSTANT OF $30 \, \text{km s}^{-1} \, \text{Mpc}^{-1}$

James G. Bartlett,[1] Alain Blanchard,[1] Joseph Silk,[2]
and Michael S. Turner[3,4]

[1] *Observatoire Astronomique de Strasbourg*
*Université Louis Pasteur*
*11, rue de l'Université*
*67 000 Strasbourg FRANCE*

[2] *Departments of Astronomy and Physics*
*and Center for Particle Astrophysics*
*University of California*
*Berkeley, CA 94720 U.S.A.*

[3] *NASA/Fermilab Astrophysics Center*
*Fermi National Accelerator Laboratory, Batavia, IL 60510-0500 U.S.A.*

[4] *Departments of Physics and of Astronomy & Astrophysics*
*Enrico Fermi Institute, The University of Chicago, Chicago,*
*IL 60637-1433 U.S.A.*

Although cosmologists have been trying to determine the value of the Hubble constant for nearly 65 years, they have only succeeded in limiting the range of possibilities: most of the current observational determinations place the Hubble constant between $50 \, \text{km s}^{-1} \, \text{Mpc}^{-1}$ and $90 \, \text{km s}^{-1} \, \text{Mpc}^{-1}$ [1]. The uncertainty is unfortunate because this fundamental parameter of cosmology determines both the distance scale and the time scale, and thereby affects almost all aspects of cosmology. Here we make the case for a Hubble constant that is even smaller than the lower bound of the accepted range, arguing on the basis of the great advantages, all theoretical in nature, of a Hubble constant of around $30\,\mathrm{km\,s^{-1}\,Mpc^{-1}}$. Those advantages are: (1) a comfortable expansion age that avoids the current age crisis; (2) a cold dark matter power spectrum whose shape is in good agreement with the observational data and (3) which predicts an abundance of clusters of galaxies in close agreement with that of x-ray selected galaxy clusters; (4) a nonbaryonic to baryonic mass ratio that is in better agreement with recent determinations based upon cluster x-ray studies. In short, such a value for the Hubble constant cures almost all the ills of the current theoretical orthodoxy, a flat Universe comprised predominantly of cold dark matter.

The hot big-bang model is enormously successful. It provides the framework for understanding the expansion of the Universe, the cosmic background radiation (CBR), and the primeval abundance of the light elements, as well as for the construction of a general picture of the formation of structure in the Universe [2]. Further, some would argue that the current orthodoxy, a flat Universe dominated by cold dark matter [3], is close to bringing it to an even higher level of success.

The latter opinion is not shared by all cosmologists [4]; many would argue that current challenges to the orthodoxy will upset it and perhaps even lead to the demise of the big-bang model itself [5]. Those challenges include a resolution of the age/Hubble constant dilemma, determination of the composition and of the quantity of dark matter, and the formulation of a coherent and detailed picture of the origin of structure in the Universe. As we shall now describe, most of these problems become successes should the Hubble constant be found to have a value of around $30\,\mathrm{km\,s^{-1}\,Mpc^{-1}}$. Further, confirmation of the cold dark matter model would provide considerable insight into the earliest moments of the Universe.

Consider first the question of the age of the Universe. In the absence of a cosmological constant, the present age and Hubble constant are related by

$$t_0 = f(\Omega_0) H_0^{-1} \simeq 9.8\,\mathrm{Gyr}\,f(\Omega_0)(H_0/100\,\mathrm{km\,s^{-1}\,Mpc^{-1}})^{-1}, \qquad (1)$$

where $f(\Omega_0)$ is a monotonically decreasing function of $\Omega_0$ ($\equiv 8\pi G\rho_{\mathrm{mean}}/3H_0^2$), obtaining a value of 1 for an empty universe ($\Omega_0 = 0$) and a value of $2/3$ for the theoretically favored flat universe ($\Omega_0 = 1$). An accurate age determina-tion is difficult, but recent values based upon dating the ages of the oldest stars are uncomfortably high: $15\,\mathrm{Gyr} \pm 3\,\mathrm{Gyr}$ [6].

If the Universe is flat and the Hubble constant is at the lower extreme of the currently accepted range, then the expansion age is barely long enough to be consistent with the age dating of the oldest stars. Even in the case of an open universe, consistency requires the Hubble constant to be at the low end of the currently accepted range. For example, the age of the Universe for a model with $\Omega_0 \approx 0.2$ and $H_0 \approx 70\,\mathrm{km\,s^{-1}\,Mpc^{-1}}$ is only about $12\,\mathrm{Gyr}$. This all becomes more severe if the oldest stars formed at modest redshifts, say $z \sim 1-2$ as might be expected in the cold dark matter model, since it would require the addition of about 3 Gyr to the age of the Universe.

There is another, less direct, indication that the age problem is a severe one, requiring a very small value of the Hubble constant. It involves "very red" galaxies observed at redshifts of order unity, e.g., the extreme case of the most distant galaxy known, with redshift $z = 4.25$ [7]. The colors of these galaxies are indicative of an old stellar population, implying substantial evolution. Such colors are hard to accommodate unless the Hubble constant is small since the age of the Universe at redshift $z$ is $t(z) = 2H_0^{-1}/3(1+z)^{3/2} \simeq 1.8\,\mathrm{Gyr}\,(30\,\mathrm{km\,s^{-1}\,Mpc^{-1}}/H_0)$ (taking $\Omega_0 = 1$ and $z = 4.25$).

Of course, a cosmological constant can ease the age problem to a degree, though at a price—the introduction of another parameter. In a flat model where the cosmological constant accounts for 80% of the mass density, $H_0 t_0 \approx 1.1$ and $t_0 \gtrsim 15\,\mathrm{Gyr}$ for $H_0 \lesssim 70\,\mathrm{km\,s^{-1}\,Mpc^{-1}}$.

Next, consider the impact of a lower value of the Hubble constant on the cold dark matter (CDM) model of structure formation. The detection of an anisotropy in the temperature of the CBR by the Differential Microwave Radiometer (DMR) on the Cosmic Background Explorer (COBE) satellite two years ago provided the first evidence for the density inhomogeneities that seeded all the structure seen today, and thus gave important confirmation of the gravitational instability theory of structure formation [8]. The cold dark matter model is the most detailed, most studied, and perhaps most successful attempt to construct a coherent picture of structure formation. However, as the quality and quantity of the data that probe the power spectrum of density inhomogeneities has improved, the case for a discrepancy between the prediction of the simplest version of CDM and the data has grown stronger (see Fig. 1).

Cold dark matter models are predicated on a flat Universe with adiabatic,

nearly scale-invariant density perturbations and matter comprised mainly of very slowly moving particle relics such as axions or neutralinos. The simplest version of CDM, where the perturbations are precisely scale invariant and the matter content consists exclusively of baryons and CDM particles, cannot simultaneously accommodate the amplitude of the fluctuations as measured by COBE, the large-scale structure as observed by galaxy surveys such as APM, QDOT, and others, the abundance of x-ray clusters, and the small-scale pairwise velocities of galaxies.

A quantitative estimate of the problems faced by standard CDM, by which we shall mean CDM with $h = 0.5$, comes from the observational power spectrum compiled by Peacock and Dodds [9]. Their work incorporates corrections for redshift-space distortions and nonlinear clustering and utilizes five different catalogues that probe inhomogeneity on length scales from $10h^{-1}\,\mathrm{Mpc} - 200h^{-1}\,\mathrm{Mpc}$ (hereafter $H_0 \equiv 100h\,\mathrm{km\,s^{-1}\,Mpc^{-1}}$). They conclude that the power spectrum of standard CDM has the wrong shape.

To be more precise, while the primeval density perturbations are scale invariant, the fact that the Universe made a transition from radiation domination to matter domination at a redshift of about $z_{\mathrm{EQ}} \simeq 2.4 \times 10^4 \Omega_0 h^2$ does impose scale on the power spectrum: $k_{\mathrm{EQ}} \simeq 0.5(\Omega_0 h^2)\,\mathrm{Mpc}^{-1}$, the scale that crossed inside the horizon at matter-radiation equality. The shape of the power spectrum seen at late epochs is determined by this scale times $H_0^{-1}$ (since observations rely on redshift as a distance indicator), leading to the definition of a shape parameter $\Gamma = \Omega_0 h \propto k_{\mathrm{EQ}} \times h^{-1}\,\mathrm{Mpc}$, which in standard CDM assumes a value of about 0.5. Peacock and Dodds [9] conclude that the data are best fit by a shape parameter $\Gamma = 0.25 \pm 0.05$.

The simplest way of achieving this is a low value for the Hubble constant! It has the additional effect of increasing the baryon fraction predicted by the theory of primordial nucleosynthesis (see below), which in turn further alters the shape of the power spectrum by suppressing power on galactic scales. To retain sufficient power on galactic scales, we favor a value of $30\,\mathrm{km\,s^{-1}\,Mpc^{-1}}$ for the Hubble constant rather than $25\,\mathrm{km\,s^{-1}\,Mpc^{-1}}$. Figure 1 compares the power spectrum obtained by Peacock and Dodds with the prediction of CDM for several values of $H_0$. (We have used the CDM power spectrum of Ref. [10], normalized to the COBE DMR fluctuation amplitude. The galaxy data is normalized to IRAS counts in spheres of $8h^{-1}\,\mathrm{Mpc}$.)

Other fixes, such as mixed dark matter, CDM + cosmological constant, raising the energy level in relativistic particles (which delays matter-radiation equality and therefore has the same effect as lowering the Hubble constant), and "tilting" the primeval power spectrum have also been proposed to address the "shape problem" [11]. The power spectrum in the mixed dark matter scenario falls dramatically on small scales and so this version of CDM has difficulty accounting for the early formation of objects such as QSOs and large, unbound groups of galaxies at high redshift. With the exception of the model containing a cosmological constant, these variants do not address the problem of the dark matter to baryonic matter ratio as measured in clusters like Coma, which we discuss below. The cosmological constant fix, however, cannot account for the large value of $\Omega_0$ inferred from the analysis of peculiar velocities in our local neighborhood [12]. At the very least, a low value for the Hubble constant is the most economical solution.

A problem for standard CDM not unrelated to the shape problem is excessive power on small scales. An often-used measure of inhomogeneity on small scales is the variance of the mass in spheres of radius $8h^{-1}$ Mpc, or $\sigma_8$; for reference, the variance of optical galaxy counts in such spheres is unity [13]. In standard, COBE normalized CDM, $\sigma_8 = 1.2$, while a Hubble constant of $30\,\mathrm{km\,s^{-1}\,Mpc^{-1}}$ results in a significantly lower value, $\sigma_8 = 0.6$. This value agrees with the variance of IRAS galaxies (see Figure 1) and implies that bright, optical galaxies are a biased tracer of mass while IRAS galaxies better trace the mass distribution.

In the probably oversimplified linear-bias scheme, the bias factor $b \equiv (\delta n_{\mathrm{GAL}}/n_{\mathrm{GAL}})/(\delta \rho/\rho) = 1/\sigma_8$ is about 1.7. This agrees with the bias found by several authors using the abundance of galaxy clusters to probe the mass fluctuations on the same scale [14]. We have used the Press-Schechter formalism [15] to calculate the distribution of x-ray emitting galaxy clusters as a function of temperature for several values of the Hubble constant. We compare these results to the data from Edge et al. [16] and Henry and Arnaud [17] in Fig. 2. For COBE-normalized CDM models, $h = 0.3$ provides an excellent fit to the data; it is all the more remarkable considering the extreme sensitivity of the cluster abundance to the Hubble constant, which is caused by the additional suppression of power on small scales due to the higher baryon abundance. The fit is better than that obtained for standard CDM normalized to the same value of $\sigma_8$ because a lower Hubble constant results in a "flatter" power spectrum on these scales (see Fig. 1).

A problem that plagues all unbiased, $\Omega_0 = 1$ models is the prediction of galaxy pairwise velocities on scales of around 1 Mpc that are several times larger than observed. While a bias of 1.7 helps significantly in reducing galaxy pairwise velocities, it does not do the whole job. However, Zurek et al. [18] argue that velocity bias, primarily due to merging, reduces the theoretically predicted velocities by about 30 percent and that observational bias, which arises in interpreting pairwise velocities in a sample contaminated by Virgo infall (and may be corrected for by treating both data and simulations identically) raises "observed" pairwise velocities by about 50 percent. In addition, Bartlett and Blanchard [19] have shown that the interpretation of the pairwise velocities is sensitive to the unknown distribution of mass around galaxies, and, on the basis of a simple model, suggest that this may alleviate the problem even further.

An often-debated issue within the context of CDM models [4], especially those with reduced power on small scales, is whether objects seen at high redshift such as quasars can indeed form sufficiently early [20]. Figure 3, adapted from Ref. [21], displays redshifts of formation for objects of various mass in COBE-normalized CDM models for different values of the Hubble constant. The formation redshift is the epoch when most of the baryons are in nonlinear objects of the specified mass. Since quasars and even massive galaxies, especially those seen at high redshift, are rare objects, one can multiply the formation redshift (plus one) by, for example, a factor of 3 for $3\sigma$ fluctuations, since the fluctuation amplitude in the linear regime grows as $(1+z)^{-1}$. We conclude that even for a Hubble constant as low as $30\,\mathrm{km\,s^{-1}\,Mpc^{-1}}$, sufficiently early formation of rare massive galaxies, the likely hosts of quasars, can occur by $z=5$, the epoch of formation of the first quasars.

The final important argument in favor of a low value for the Hubble constant comes from primordial nucleosynthesis and recent determinations of the ratio of dark-to-baryonic matter in rich, x-ray emitting clusters such as Coma. With a single parameter, the baryon-to-photon ratio $\eta$, primordial nucleosynthesis successfully accounts for the abundances of the four lightest elements D, $^3$He, $^4$He, and $^7$Li, provided that $\eta \simeq (2.5-6)\times 10^{-10}$. This leads to the best determination of the baryon density, $\rho_B = (1.7-4.1)\times 10^{-31}\,\mathrm{g\,cm^{-3}}$ [22]. However, because the critical density depends upon the Hubble constant, $\rho_{\mathrm{crit}} = 3H_0^2/8\pi G$, the fractional contribution of baryons to the critical density also depends upon the Hubble constant,

$$\Omega_B \equiv \rho_B/\rho_{\mathrm{crit}} \simeq 0.01 h^{-2} - 0.02 h^{-2}. \qquad (2)$$

The fractional contribution of baryons increases with a lower Hubble constant, though still must be less than about 0.2 even if $h = 0.3$. (An even more extreme view than ours has long been advocated by Shanks [23], who has argued for a value of $H_0$ as low as $30\,\mathrm{km\,s^{-1}\,Mpc^{-1}}$ in order to revive a baryon-dominated Universe. However, both primordial nucleosynthesis and recent detections of CBR anisotropy on various angular scales are inconsistent with $\Omega_B \sim 1$ [24].)

An accurate determination of $\Omega_B$ leads to a test of the orthodoxy that has been much emphasized recently: the ratio of total-to-baryonic mass in a system large enough to represent a fair sample of the Universe should be $\Omega_B^{-1} \simeq 50h^2 - 100h^2$. Briel et al. [25] have pointed out that the baryon-to-dark matter ratio in clusters could be problematic. Based on their data, White et al. [26] have concluded that the ratio of total mass to mass in x-raying emitting gas is about $(20 \pm 5)h^{3/2}$. (Essentially all the "visible" mass in baryons is in x-ray emitting gas; in this analysis it is assumed that the dark matter in clusters is *not* comprised of dark baryons.)

No value of $H_0$ within the traditionally accepted range can account for this observation (for $\Omega_0 = 1$). For $H_0 = 30\,\mathrm{km\,s^{-1}\,Mpc^{-1}}$, the orthodoxy is consistent with the data, but only just; the measured ratio differs from the nucleosynthesis prediction by about two standard deviations. However, it is likely that systematic effects still remain, most of which go in the direction of increasing the total-to-baryonic mass ratio.

For example, the baryon-to-dark matter ratio is likely to be somewhat enhanced due to the settling of baryons [26]. Mapping of the mass distribution in clusters by studying the shear of background galaxy images produced by gravitational lensing results in an estimated mass that exceeds that obtained from application of the virial theorem to the hot gas, in two separate cases by a factor of about three [27]. This result makes sense if the clusters are not in virial equilibrium, or if the gas is partially supported by magnetic fields. The former possibility is inferred to be the case for the hot, x-ray emitting gas that is still undergoing infall according to cluster simulations and is also seen to show substructure [28]. That this may be a more or less ubiquitous phenomenon is suggested indirectly by the requirement that substantial amounts of intracluster gas must have only merged recently, moving at a speed comparable to the sound velocity, in order to provide enough ram pressure to account for radio source morphologies [29]. Ensuing gas clumpiness would also lower the inferred gas mass, and together with upward correction of cluster mass estimates, could comfortably reconcile cluster gas content with nucleosynthesis predictions for $H_0 = 30 \,\mathrm{km\,s^{-1}\,Mpc^{-1}}$.

While our arguments for a low value of the Hubble constant have revolved around the CDM cosmogony, even in the simplest cosmological model, one where baryons constitute the sole form of matter, the value of the baryon density given by nucleosynthesis is uncomfortably small compared to many estimates of $\Omega_0$, unless the Hubble constant is small. In particular, assuming that the mass-to-light ratios derived for clusters are representative of the Universe as a whole, one can derive $\Omega_0$ from the mean luminosity density. Such measurements indicate that $\Omega_0 \simeq 0.1 - 0.4$ [30]. In a pure baryonic model, the Hubble constant has to be smaller than about $32 \,\mathrm{km\,s^{-1}\,Mpc^{-1}}$ to accommodate $\Omega_0 = \Omega_B \gtrsim 0.2$. Moreover, if the recent *tentative* detection of deuterium absorption at the level $\mathrm{D/H} \sim 2 \times 10^{-4}$ in an intergalactic cloud towards a quasar at redshift $z = 3.32$ proves correct [31], then the problem is even more acute. To produce this much deuterium, $\Omega_B h^2$ must be around 0.005, which requires a Hubble constant of $22 \,\mathrm{km\,s^{-1}\,Mpc^{-1}}$ just to achieve $\Omega_B \simeq 0.1$.

Finally, let us turn to the testing of our provocative suggestion. In the near term, the best prospects for *indirect* confirmation involve measurements of CBR anisotropy [32]. Compared to standard CDM, degree-scale CBR anisotropy is predicted to be about a factor of 1.5 larger (see Fig. 4). While the experimental situation on the degree scale is not settled at the moment, there is statistical support for the higher range of detections, such as those of the MAX [33] and Python [34] collaborations [35]. These detections are compatible with CDM, provided $H_0 \lesssim 50 \,\mathrm{km\,s^{-1}\,Mpc^{-1}}$.

The ultimate test of course is to measure the Hubble constant itself. It has recently been argued that a variety of techniques provide convincing evidence for a Hubble constant of $80 \pm 10 \,\mathrm{km\,s^{-1}\,Mpc^{-1}}$ and that the question will soon be settled when Cepheid variables are studied by the refurbished Hubble Space Telescope [1]. However, most of the techniques that are converging on this value, including Tully-Fisher, surface-brightness fluctuation, fundamental-plane, and planetary-nebulae techniques, involve the same lower rungs on the infamous distance ladder—and thus could have a common systematic error. From our perspective, the most troublesome measurements are those based upon type II (core collapse) supernovae; they "jump" the distance ladder—and thus do not share common systematic errors with the previously mentioned methods—and still give a value consistent with $80 \pm 10\,\mathrm{km\,s^{-1}\,Mpc^{-1}}$ [36].

There are two lines of defense for our hypothesis: (1) common systematic error in the empirically based determinations and an error in the type II supernovae determination; or (2) current measurements have yet to reach sufficient distances to sample the Hubble flow (or are still influenced by Malmquist bias). Both possibilities have some merit [37].

The paucity of nearby giant elliptical galaxies means that the calibration of the of fundamental-plane, surface-brightness fluctuation, and planetary-nebulae techniques are entirely based upon the bulges of M31 and M32, and thus inadequate. Intermediate-age populations are found to be present in nearby examples of bulges and dwarf ellipticals [38], and their effect on these distance indicators is unexplored.

The Tully-Fisher technique has been criticized on the grounds that its local spiral calibrators are excessively blue (in B-I band) for their linewidths. If this is a stellar-population effect, it may be circumvented by working in I band or in the near infrared. However, two type Ia supernovae, which traditionally favor a low value of the Hubble constant, have been recently calibrated with Cepheids in Tully-Fisher calibrator galaxies. The calibration disagrees with the I-band Tully-Fisher distances, and to save the latter, this has been attributed to dust absorption in the vicinity of the supernovae [39]. This tends to raise the luminosities of the nearby supernovae used as calibrators, relative to distant, presumed dust-free supernovae, thereby bringing the two techniques into agreement for a larger value of $H_0$. However if the nearby type Ia supernovae calibrators are intrinsically underluminous, as seems to be the case for at least half the type Ia's [40], there could be a possible bias towards a lower global value of $H_0$. This would leave one with a discrepancy in the Tully-Fisher distance indicators.

It is also worth considering the possibility that our local volume of the Universe may not be representative, which is reasonable as large-scale variations are expected in any CDM model. A "large" local value for $H_0$ would occur if we lived in an underdense region. Simulations have shown that a fifty percent reduction in $H_0$ is possible if measurements are performed within about $20h^{-1}\,\mathrm{Mpc}$; specifically, the reduction is $90(-\Delta n_{\mathrm{gal}})\sigma_8\Omega^{0.6}$ per cent [41], where $-\Delta n_{\mathrm{gal}}$ is the galaxy number underdensity. A 70 percent underdensity would reduce $H_0$ by 40 percent for $\sigma_8 = 0.7$; such fluctuations are expected out to about $20h^{-1}\,\mathrm{Mpc}$, or 4 correlation lengths, if one is uti-lizing luminous galaxies which are $\sim 3\sigma$ fluctuations. Such a local "hole" is also consistent with, and would even account for, the apparent evolution in number seen in APM galaxy counts [42] at an otherwise surprisingly low redshift ($z \sim 0.2$). Finally, we note that the only truly global measurements of $H_0$ (i.e., measurements at $z \gtrsim 0.1 - 0.2$), namely those utilizing the Sunyaev-Zel'dovich effect in galaxy clusters and time-delay measurements in the images produced by the gravitational lensing of a variable quasar, favor a value of $H_0$ systematically lower than is obtained from the more local measurements. A recent study of A2218 using the Interferometric Ryle Telescope places $H_0$ in the range 20–75 $\mathrm{km\,s^{-1}\,Mpc^{-1}}$ [43], while modeling of the double quasar Q0957+561 gives $H_0 = 37 \pm 14$, if one adopts the time delay of Press et al. [44] and the velocity dispersion of Rhee [45].

All this being said, reconciling $H_0 = 30\,\mathrm{km\,s^{-1}\,Mpc^{-1}}$ with the bulk of the measurements of the Hubble constant is not easy. In fact, our strongest argument for a significant systematic error in the distance scale is that history is on our side! The techniques that both jump the distance ladder and provide a measurement of the global Hubble constant (Sunyaev-Zel'dovich effect and time delay in gravitational lens systems) should eventually provide a definitive answer. Because of the many advantages of $H_0 = 30\,\mathrm{km\,s^{-1}\,Mpc^{-1}}$ spelled out here, we believe that it is worth keeping an open mind.

## Acknowledgments

We are very grateful to J. Peacock for kindly providing us with the power spectrum data presented in Fig. 1, to M. Tegmark for preparing Fig. 3, and to N. Sugiyama for preparing Fig. 4. This work was supported in part by the DOE (at Chicago and Fermilab), by the NASA through grant NAGW-2381 (at Fermilab), by the NSF and DOE at Berkeley, and by a Chateaubriand fellowship from the Centre International des Étudiants et Stagiaires (Strasbourg).

# Figure Captions

**Figure 1:** CDM power spectra vs. observations, as compiled by Peacock and Dodds [9]. The solid lines show the CDM power spectrum [10], normalized to the COBE amplitude ($Q_{\rm rms} = 17\mu{\rm K}$) for the indicated values of the Hubble constant and accounting for the suppression caused by an abundance of baryons consistent with primordial nucleosynthesis. The dashed lines show the CDM power spectrum for $h = 0.2, 0.3$ with $\Omega_B = 0$, showing the importance of the suppression of power on small scales for small values of $h$.

**Figure 2:** The distribution of x-ray emitting galaxy clusters as a function of temperature. The curves represent the number predicted in the Press-Schechter formalism [15] for several values of $h$. All of the underlying CDM power spectra have been COBE normalized and include the effect of suppression of power on small scales due to baryons. The data come from Edge et al. [16] and Henry and Arnaud [17]. The dashed line highlights $h = 0.3$ which provides a remarkably good fit to the data.

**Figure 3:** Virialization redshifts for objects of various mass. The virialization redshift is defined to be the redshift at which the bulk of the matter condenses into objects of a specified mass (baryonic + CDM) multiplied by the bias factor (approximately 1.7; see text). Details of the calculation are given in Ref. [21]; this figure was prepared by M. Tegmark.

**Figure 4:** Angular-power spectrum for standard CDM with a Hubble constant of $30, 50, 75\,{\rm km\,s}^{-1}\,{\rm Mpc}^{-1}$ (solid, dotted, dashed lines respectively), with $\Omega_B h^2 = 0.015$. Calculations provided by N. Sugiyama; compilation of anisotropy experiments supplied by D. Scott and M. White [35]. The CBR temperature anisotropy predicted for a given experiment depends upon its filter function; very roughly, for an experiment that measures the temperature difference on angular scale $\theta$, $\delta T/T \sim \sqrt{l(l+1)\langle|C_l|^2\rangle}$ for $l \sim 200°/\theta$. The horizontal error bars represent the half-peak points of the window functions; the precise location of the vertical error bars is spectrum-dependent, and indicative values only are shown. Error bars are $\pm 1\sigma$; upper limits are $2\sigma$.